\documentclass[useAMS]{mn2e}
\usepackage{epsfig}
\title[RCB stars and X-ray sources]{Strong [O III] and [N II]
emission lines in globular clusters from photoionized R Corona
Borealis star winds}

\author[Maccarone \& Warner]{Thomas J. Maccarone \& Brian
Warner\thanks{also Department of Astronomy, University of Cape
Town. Cape Town, South Africa} \\ School of Physics and Astronomy,
University of Southampton, Hampshire SO17 1BJ,United Kingdom\\ }

\begin{document}
\def\ltsim{\mathrel{\rlap{\lower 3pt\hbox{$\sim$}}
        \raise 2.0pt\hbox{$<$}}}
\def\gtsim{\mathrel{\rlap{\lower 3pt\hbox{$\sim$}}
        \raise 2.0pt\hbox{$>$}}}
\def\1399{CXOJ033831.8 - 352604}

\date{}

\pagerange{\pageref{firstpage}--\pageref{lastpage}} \pubyear{}

\maketitle

\label{firstpage}

\begin{abstract}
The globular cluster X-ray source CXO~J033831.8-352604 in NGC~1399 has
recently been found to show strong emission lines of [O III] and
[N II] in its optical spectrum in addition to ultraluminous X-ray
emission with a soft X-ray spectrum.  It was further suggested that
this system contained an intermediate mass black hole which had
tidally disrupted a white dwarf, producing the strong emission lines
without detectable hydrogen emission.  We show that an alternative
exists which can explain the data more naturally in which the oxygen
and nitrogen rich material is ejecta from a RCB star, or a tidal
disruption of an RCB star or a hydrogen-deficient carbon star.
The scenario we propose here does not require an intermediate mass
black hole as the accretor, but also does not exclude the possibility.
\end{abstract}


\section{Introduction}

The discovery of globular cluster X-ray sources in the Galaxy
(e.g. Clark 1975) in the mid-1970's prompted the suggestion that X-ray
emission might come from accretion onto intermediate mass black holes
(Bahcall \& Ostriker 1975 -- BA75; Silk \& Arons 1975 -- SA75).  It
has since been shown that all the bright Galactic X-ray sources are
likely to have neutron star accretors -- all but one of them show
pulsations and/or surface thermonuclear runaways (Liu et al. 2007 and
references within; see also Altamirano et al 2010).  The other has
been the subject of Doppler tomography which favours a neutron star
accretor, and rules out an intermediate mass black hole accretor (van
Zyl et al. 2004).  The mechanism suggested by BO75 and SA75 is
nonetheless still potentially relevant for extragalactic globular
cluster X-ray sources, as well as for faint X-ray (Ho et al. 2003) and
radio (Maccarone 2004) sources in Galactic globular clusters.  Recent
work has suggested fueling of such black holes through tidal
destruction of stars, which may be an effective way to provide more
material in the vicinity of the black hole than standard stellar mass
loss provides (e.g. Rosswog et al. 2009).

It has long been realized that the large globular cluster populations
of nearby giant elliptical galaxies made them potentially rich targets
for searching for globular cluster X-ray sources (Fabian, Pringle \&
Rees 1976).  The {\it Einstein} observatory had the sensitivity and
angular resolution to detect bright non-nuclear point sources in
nearby galaxies, but only with the launch of {\it Chandra} has it been
possible to localize such sources well enough to make reliable
associations between the X-ray sources and globular clusters at
distances beyond that of M~31.  In recent years, several strong
candidates for black holes in globular clusters have been identified
(Maccarone et al. 2007; Brassington et al. 2010; Irwin et al. 2010;
Shih et al. 2010; Maccarone et al. 2010).  Two particularly
interesting globular cluster X-ray sources show strong optical
emission lines.  The first, in NGC~4472, shows highly variable
emission, peaking at about $4\times10^{39}$ ergs/sec, confirming that
the bulk of the emission comes from a single X-ray source (Maccarone
et al. 2007), and strong, broad [O III] lines, with the lack of a
clear detection of H$\beta$ emission implying a ratio of oxygen to
hydrogen much larger than that of solar composition material (Zepf et
al. 2007,8; Steele et al. 2010, submitted to ApJ).  The second, \1399
in the Fornax xluster galaxy NGC~1399 is a bit fainter
($L_X$$\approx$$2\times10^{39}$ ergs/sec), and has not shown strong
variability, but shows strong emission lines from both [O III] and [N
II] -- although these lines are substantially fainter and narrower
than the lines seen from RZ~2109 (Irwin et al. 2010 -- I10).  I10
argued that the properties of this source could be explained by tidal
disruption of a white dwarf by an intermediate mass black hole.  In
this letter, we discuss a new interpretation for this system -- that
the globular cluster contains an RCB star, the wind of which is
photoionized by the X-ray source.

\section{Data}

We present no new data in this paper, but do review the observational
findings of I10 for the benefit of the reader.  They report three
X-ray observations with at least 100 counts, all showing an X-ray
luminosity of $1.5-2.3\times10^{39}$ ergs/sec and good spectral fits
with models of power laws with $\Gamma$ in the 2.5-3.0 range, or disc
blackbodies (Mitsuda et al. 1984) with $KT_{in}$ of 0.36-0.39 keV.

They made several optical spectra.  They find emission lines of the [O
III]$\lambda$ 5007 doublet and the [N II]~$\lambda$~6584~\AA\
doublet.  They report HWHM of the lines to be 70 km/sec and their
plots show a peak flux density in [O III]$\lambda$~5007~\AA\ of
$1.7\times10^{-18}$ ergs/sec/cm$^2$/\AA\ corresponding to a luminosity
of about $2.0\times10^{35}$ ergs/sec, given the 70 km/sec HWHM and in
[N II]~6584~\AA\ of $2.0\times10^{-18}$ ergs/sec/cm$^2$/\AA,
corresponding the a luminosity of about $2.5\times10^{35}$ergs/sec.
No other lines are found in this spectrum (J. Irwin, private
communication).  Visual inspection of their plots shows a noise level
of about $3\times10^{-19}$ ergs/sec/cm$^2$/\AA, so we can set as
targets for photoionization calculations that the luminosity in the
main peaks of [O III] and [N II] are roughly similar, and both at
about $2\times10^{35}$ ergs/sec, and that any other lines in the
optical bandpass should be no more than about half the strength of
those two lines.

\section{The problem with the disrupted white dwarf scenario}
The key problem with the intepretation of I10, which they noted, but
did not resolve, is that the nitrogen emission lines are stronger than
the oxygen emission lines.  Carbon-oxygen white dwarfs rarely show
nitrogen emission or absorption lines; in the process of producing a
carbon-oxygen (CO) white dwarf, the core nitrogen is usually consumed
almost entirely in the helium burning phase.  Helium white dwarfs can
contain substantial nitrogen, but helium white dwarfs are low mass
objects formed through a binary evolutionary pathway that causes the
atmosphere of a low mass evolved star to be ejected or transferred
before the nuclear burning runs its course.  As low mass objects,
extreme fine tuning of the impact parameter of their interactions with
an intermediate mass black hole is needed to allow them to be tidally
disrupted, rather than tidally detonated (Rosswog et al. 2009). While
no helium lines were found in the source spectrum, we show below that
this is not a serious constraint on the amount of helium present, as
there are plausible ranges for plasma temperature and ionization
parameter for which helium can be the dominant species, but no helium
lines will be detected.

\section{An alternative explanation: RCB stars}

There are three closely related classes of stars which are nearly
hydrogen-free and which have much larger ratios of nitrogen to oxygen
than do white dwarfs -- R Corona Borealis (RCB) stars, extreme helium
(EHe) stars and hydrogen-deficient carbon (HdC) stars (although we
note that some authors consider the RCB stars to be a subclass of the
HdC stars -- e.g. Warner 1967).  The prevailing formation mechanism
for these classes is the merger of a He WD with a CO WD (e.g.  Webbink
1984; Warner 1967; Clayton et al. 2007; Garcia-Hernandez et al. 2009),
but there are alternative suggestions that single star evolution can
produce a ``final flash'' of helium burning (Iben et al. 1996).  Since
the formation of RCB stars relies on a process related to close binary
evolution, and the production of He white dwarfs can be enhanced by
stellar collisions, one might expect an overabundance of RCB stars and
HdC stars in globular clusters, in the same way that X-ray binaries
are overabundant in dense globular clusters compared with field star
populations.  There is some evidence that globular clusters contain
substantial populations of stars with strongly enhanced carbon in
their atmospheres (e.g. Strom \& Strom 1971; Zinn 1973), but such
studies are anecdotal at the present time.  The abundance patterns in
those stars are generally attributed to dredge-up, rather than
mergers, but binary evolutionary processes may be necessary
(e.g. Lucatello et al. 2005).  We are not aware of any systematic
search for RCB, EHe or HdC stars in globular clusters -- although as
discussed below in section 4.3, some of the surveys for variable stars
in globular clusters would have turned up some of the known RCB stars
(see e.g. Clement et al. 2001).

\subsection{Photoionization calculation}

We use the XSTAR version 2.2 photoionization package (Kallman \&
Bautista 2001) to test whether feasible parameter values can yield
reasonable emission line spectra.  We start from the assumption that
an RCB star is located somewhere in the globular cluster core, and its
wind is being photoionized by the bright X-ray source.  There are many
free parameters in play for a system of this kind, so we apply a
trial-and-error approach to find a plausible set of parameters that
reproduces the observed optical spectrum to within a factor of a few.
Because many of the parameters are nearly degenerate with one another
and, apart from upper limits we have only two constraints (since the
ratio of the strengths of the lines within the doublets is fixed by
atomic physics), we can state with near certainty that the parameter
values here are {\it not} unique.

We start from the assumption that the photoionized gas will have
chemical composition roughly similar to that of the RCB stars (see
e.g. Garcia-Hernandez et al. 2009).  We set the abundances in XSTAR
for helium to 4 times the solar abundance, for carbon to 10 times the
solar abundance, nfor itrogen to 30 times the solar abundance and for
oxygen to 5 times the solar abundance.  We note that the study of
Garcia-Hernandez allowed for a large range of oxygen and helium
abundance but a small range for the other parameters.  We leave out
the lines from the heavier elements because they are relatively low in
abundance and are not reported to have been seen, and because XSTAR
calculations are sped up significantly by setting some abundances to
zero.  We set the temperature of the plasma to 7500 K (a typical
temperature for warm RCB stars), then density to $10^3$ particles
cm$^{-3}$, the ionizing source to be a $4\times10^6$ K blackbody with
luminosity $2\times10^{39}$ ergs/sec, the ionization parameter $\Xi$
to $10^{0.5}$, and the column density to $2\times10^{19}$ cm$^{-2}$.
The emission in this model comes from a thin shell at a radius of
$8\times10^{17}$ cm from the X-ray source.  For this set of
parameters, we find $L_{6584}=2.6\times10^{35}$ ergs/sec, $L_{5007}$
is 1.1$\times10^{35}$ ergs/sec, and the strongest optical line from a
species not reported by I10 is the He~I~$\lambda$~4686 line, with
$L=4\times10^{34}$, well below the detection threshold in the spectra
shown by I10.  The calculations are thus in reasonably good agreement
with the observed data, but the thinness of the shell, about
$2\times10^{16}$ cm is difficult to explain in a physically viable
scenario.

We can alternatively consider the case where a stellar wind is being
ionized.  In this case, the wind is likely to subtend only a fraction
of the solid angle seen from the accretor.  XSTAR assumes spherical
symmetry, so the calculation done will have to be a crude
approaximation of the realistic geometry.  Larger column densities are
necessary in this case, but also, the large column densities can exist
without affecting the X-ray spectrum of the source, since the column
need not be between the observer and the X-ray source.  The region
where $\delta R\sim R$ should contribute most strongly to the observed
emission.  The inner regions will have small solid angles
contributing, and the outer regions will have low densities due to the
$R^{-2}$ dependence of stellar wind density.  We run XSTAR again, with
a density of 400 particles cm$^{-3}$, column density to
$3\times10^{20}$ cm$^{-2}$, and log $\Xi$=0.75, and all other
parameter values as above.  The range of radii in the calculations
then extends from $9\times10^{17}$ to $1.7\times10^{18}$ cm$^{-2}$.
This calculation yields line luminosities of $5\times10^{35}$ and
$3\times10^{35}$ ergs/sec, for the brighter lines of [N II] and [O
III], respectively, after accounting for the fact that only about 10\%
of the solid angle on the sky is emitting.  We can then compute the
mass loss rate expected from the wind of an RCB star if its density is
400 cm$^{-3}$ a distance $4\times10^{17}$ cm out, and find that it
will be about $10^{-5}$ $M_\odot$ per year, towards the upper end of
the range observed from these objects.  The gas should reach this
radius on a timescale of a few thousand years, again, well within
reasonable values for the lifetimes of these stars.  Given that an
acceptable solution has been found, and that the degeneracies allow
other parameter values, it is clear that if RCB stars are
abundant enough in globular clusters, it is plausible for the observed
optical emission lines to come from photoionization of an RCB wind.
We also note that the winds from RCB stars typically are $\sim100-200$
km/sec in velocity (e.g. Clayton et al. 1994; Clayton et al. 2003),
only slightly faster than the HWHM reported in I10.

The winds from RCB stars are probably driven by radiation pressure on
dust (see e.g. Clayton et al. 2003), which then drags the gas along.
This might then modify the expected photoionization signatures
expected. We do not think this is a likely scenario, since we have
assumed a temperature of 7500 K for the gas -- a temperature at which
the dust would be sublimated before travelling the $\sim$ light year
out to the region in which the photoionized emission lines are
produced.

\subsection{The black hole mass of the photoionzing source}

The implications for the mass of the black hole doing the
photoionizing are weak.  X-ray sources at luminosities of
$2\times10^{39}$ ergs/sec in the Milky Way are often found to be in
the ``very high'' spectral state, in which their spectra can be
dominated by a $\Gamma\approx2.7$ power law, in good agreement with
the spectrum presented by I10.  On the other hand, if one takes the
best disc blackbody fit from I10, one finds that the inner disc radius
should be about 1500 km, assuming a colour correction factor of about
3 (see e.g. Davis et al. 2005), and assuming the disc is observed
face-on.  This would correspond to a black hole of $\sim100-1000$
$M_\odot$, depending on the inclination angle -- the lower end of the
range would imply a Schwarzschild black hole observed pole-on, while
the upper end of the range would imply a Kerr black hole viewed close
to edge-on.  The upper end of the range is disfavored by the fact that
stellar mass black holes (Maccarone 2003) and active galactic nuclei
(Ho 1999; Maccarone, Gallo \& Fender 2003) tend to have hard power law
spectra below a few percent of their Eddington luminosities.  The soft
X-ray spectrum is thus not a diagnostic of whether the accretor is of
stellar or intermediate mass.

We also consider the possibility that the RCB star ejecta could be the
mass supply to an intermediate mass black hole, so that no additional
bright X-ray source would need to be produced in the cluster.  Gas
within $GM/(v_w^2+c_s^2)$ can be accreted onto the black hole, in a
manner similar to Bondi accretion.  This yields $R_{acc} = 10^{15}
(M/300 M_\odot)$ cm.  At that radius, the wind density will be
$\sim10$ cm$^{-3}$ at this radius.  The Bondi rate will then be
$\sim10^{14}$ g/sec -- far too low to account for the observed X-ray
luminosity, even before accounting for the fact that the Bondi rate
seems to overestimate observed luminosities by factors of $10-100$
(Perna et al. 2003; Pellegrini 2005).

\subsection{Observable tests of the RCB hypothesis}
A potential key diagnostic is the ratio of $^{18}$O to $^{16}$O.  As
predicted by Warner (1967), the hydrogen-deficient carbon stars tend
to have about 2-3 times as much $^{18}$O as $^{16}$O (Garcia-Hernandez
et al. 2009).  Most of the RCB stars show 3-20 times as much
$^{16}$O as $^{18}$O (Clayton et al. 2007; Garcia-Hernandez et
al. 2009) -- still well above the solar value (e.g. Collier et
al. 1998).  Oxygen isotopic abundances in stars are usually measured
from molecular bands, but this is not possible for the case of \1399.
It is unlikely that the isotopic shift will be measurable in the 5007
line itself, especially if $^{16}$O is the dominant species, and
merely not as dominant as it normally is.  We were not able to find
any calculations in the literature for the magnitude of the isotopic
shift for [O III]$\lambda$ 5007, but were able to find that the shift
between $^{11}B$ III and $^{10}$B III is about 1 part in 50000 (Litzen
\& Kling 1998).  The fractional mass difference for $^{18}$O versus
$^{16}$O is slightly larger, but the B III transition takes place
closer to the nucleus.  The magnitude of the wavelength shift is
likely to be too small to see for the [O III]$\lambda$ 5007 given the
broadening of 70 km/sec reported by I10, but without a careful
calculation of the wavelength shift expected for the isotopic
difference, it is not clear whether a precise centroiding of the [O
III] might be useful -- the centroiding of the line in I10 is likely
to be accurate to only about 5 km/sec -- probably larger than the
isotopic shift, especially after weighting by the expected
$^{16}$O/$^{18}$O.

A few other strong lines are expected, based on our XSTAR simulations,
but outside the optical bandpass used by I10.  The strongest line
should be the 26 $\mu$m [O IV] line, and the strongest line observable
with a ground-based CCD should be the 10830~\AA\ line of He I, which
should have a flux about 10 times lower than those of the stronger
components of the [O III] and [N II] doublets (but only a factor of
about two weaker than the [O III]~$\lambda$~4959~\AA line).  The [N
III] line at 57 $\mu$m is one of the strongest lines from the source
-- just weaker than [N II], and just stronger than [O III], but just
falls within the bandpass of Herschel, and is about 1000 times too
weak for Herschel to detect.  A strong O VII line at 22~\AA\ should
also be present, but again, will have a line luminosity well below the
detection thresholds for existing instruments in the X-rays.  It thus
may be possible, with very long integrations, to detect He I, but the
main prediction of the model is that other spectral lines will be very
hard to detect.

The idea has an alternative possible test -- whether there is a
sufficiently large population of RCB stars in the Galactic globular
cluster population.  It has been estimated that the total Galactic
population of RCB stars is about 3200 by scaling up the much better
constrained LMC population size by the ratio of stellar masses (Alcock
et al. 2001).  Since only about 0.1\% of the Galactic stellar mass is
in globular clusters, one would then find that there should be only a
few RCB stars in the entire Galactic globular cluster population.
However, if the double degenerate scenario for producing these stars
is the correct one, then one should expect a substantial dynamical
enhancement in the numbers of RCB stars in globular clusters relative
to field star populations.  Using the canonical factor of 100 that
applies to X-ray binaries, a substantial fraction of the stars in
globular clusters that appear to be on the asymptotic giant branch
should actually be RCB stars.  The recent finding that RCB stars
separate themselves from standard AGB star populations very well in
the mid-infrared (Tisserand et al. 2010), due to increased dust
re-emission, should make searches for them in the cores of globular
clusters feasible in the near future using ground-based systems at 10
$\mu$m.  On the other hand, even a dynamical enhancement of a factor
of a few in the fraction of cluster stars that are RCB stars relative
to the same fraction for the field would probably be enough to make
our scenario plausible -- the ``extra'' RCB stars should be
concentrated in cluster cores where the bulk of stellar interactions
take place, and should additionally be concentrated in the most
massive clusters, since these tend to have the highest stellar
interaction rates (e.g. Smits et al. 2006).  In the context of our
proposed scenario, it will be possible to make an estimate of the
dynamical enhancement of RCB stars in globular clusters once a large
number of spectra of globular clusters with bright X-ray sources have
been published.  At the present time, the sample of such objects is
small, and possibly biased towards the clusters with emission lines.

We do note that there have been surveys for variable stars in globular
clusters, and that some of these surveys have been made over time
baselines of a decade or more (e.g. Sawyer Hogg 1980; Clement et
al. 2001).  These surveys make it clear that there is not a large
population of RCB stars which fade as frequently as the RCB stars
which have been discovered to date.  Given the diversity of RCB stars
in terms of how frequently they show fading events, however, it is not
clear whether there is a sizeable population of RCB stars whose fading
events are infrequent enough to have been missed so far.  There are
RCB stars which have shown only one fading event over long durations
(e.g. XX Cam -- one in over 100 years; UV Cas -- one in over 70 years
and Y Mus -- one in over 40 years -- see Jurcsik 1996), suggesting
that there may very well be some which show very infrequent fading
events, and that the typical timescales on which fading events happen
for the well know RCB stars are significant underestimates of the
typical intervals between fading events for the population of RCB
stars as a whole.  Additionally, the OGLE lightcurves of known RCB
stars often vary by a tenth of a magnitude or less over extended
periods between fading events (e.g. Tisserand et al. 2010), which
could lead to misclassification as non-variable stars or
irregular/long period variables in past surveys of globular clusters
for variable stars.

Regardless, the scenario we propose does not require an overabundance
of RCB stars in globular clusters of the same factor of 100 found for
X-ray binaries.  In fact, if the overabundance were that large, we
would expect to find spectral lines like those reported by I10 in a
very large fraction of globular clusters with bright X-ray sources.
At the present time, such studies are largely anecdotal, but only one
cluster has shown such lines.  Our scenario merely requires that there
be some overabundance of RCB stars formed through dynamical
interactions, and that these stars remain in the cores of their
clusters after formation.  The factor of enhancement required will
become clear only after a large sample of spectra of clusters with
$L_X\sim10^{39}$ ergs/sec X-ray sources has been made, and the results
of those searches (including non-detections) are reported.

\section{Conclusions}

We have shown that the combination of strong nitrogen and oxygen
emission lines from \1399, coupled with a lack of hydrogen and helium
emission lines can be well explained if the cluster contains a bright
X-ray source near its center photoionizing the wind of an RCB star.
The scenario makes no requirements on the mass of the central black
hole, and the past X-ray spectral information is inconclusive on this
point -- thus the case for an intermediate mass black hole previously
made is substantially weakened by this new possibility.  If the
scenario proposed here is correct, then it is highly likely that the
formation rate of RCB stars in globular clusters is at least
moderately dynamically enhanced.

\section{Acknowledgments}
TJM thanks the European Union for support under FP7 grant 215212:
Black Hole Universe.  We thank the referee Geoffrey Clayton for a
constructive and helpful report which improved the quality of the
paper, and Jimmy Irwin for sharing unpublished details about the
spectrum of CXO~J033831.8-352604.  BW's research is supported by the
National Research Foundation of South Africa and the University of
Cape Town.

\label{lastpage}

\end{document}